\documentclass[twocolumn,fleqn,showpacs,preprintnumbers,amsmath,amssymb]{revtex4}
\def\Vec#1{\mbox{\boldmath $#1$}}
\def\upcite#1{$\,$\cite{#1}}

\usepackage{graphicx}
\usepackage{dcolumn}
\usepackage{bm}

\def\av{{\mathrm{av}}}
\def\eff{{\mathrm{eff}}}
\def\erf{{\mathrm{erf}}}
\def\Tr{{\mathrm{Tr}}}
\def\e{{\mathrm{e}}}
\def\kg{{\mathrm{kg}}}
\def\m{{\mathrm{m}}}

\def\cm{{\mathrm{cm}}}
\def\J{{\mathrm{J}}}
\def\K{{\mathrm{K}}}

\begin{document}

\preprint{APS/123-QED}

\title{Bose-Einstein condensation and superfluidity of dilute Bose gas in a random
 potential}

\author{Michikazu Kobayashi}
\author{Makoto Tsubota}%
\affiliation{Department of Physics,Osaka City University, Sumiyoshi-Ku,
Osaka 558-8585, Japan}%

\date{\today}

\begin{abstract}
There is a growing interest in the relation between Bose-Einstein condensation (BEC) and
 superfluidity.  
A Bose system confined in random media such as porous glass is suitable for studying this relation
 because BEC and superfluidity can be suppressed and controlled in such a disordered environment.  
However, it is not clear how this relation is affected by disorder and there are few theoretical
 studies that can be quantitatively tested by experiment.  
In this work, we develop the dilute Bose gas model with a random potential that takes into account
 the pore size dependence of porous glass.  
Then we compare our model with the measured low-temperature specific heat, condensate density, and
 the superfluid density of $^4$He in Vycor glass.  
This comparison uses no free parameters.  
We predict phenomena at low temperatures that have not yet been observed.  
First, the random potential causes a $T$-linear specific heat instead of the $T^3$ dependence that
 is usually caused by phonons.  
Second, the BEC can remain even when the superfluidity disappears at low densities.  
And third, the system makes a reentrant transition at low densities; that is, the superfluid phase
 changes to the normal phase again as the temperature is reduced.  
This reentrant transition is more likely to be observed when the strength of the random potential
 is increased.  
\end{abstract}

\maketitle

\section{Introduction}

Bose-Einstein condensation (BEC) and superfluidity of liquid $^4$He in random environments
 including Aerogel and Vycor glass are active problems in quantum fluid research.  
In particular, finding out how spatial confinement affects the Bose fluid has stimulated  both
 experimental and theoretical studies.  

Below the $\lambda$ temperature of 2.17 K, liquid $^4$He enters the superfluid state and behaves as
 though it has no viscosity.  
Superfluidity is a macroscopic quantum phenomenon as well as superconductivity, and understanding
 both have been one of the major goals of quantum statistical physics.  
The various observations of superfluidity was successfully explained by the phenomenological two
 fluid model\upcite{Tilley}, which is based on the idea that the system consists of an inviscid
 superfluid and a viscous normal fluid.  
On the other hand, the lambda transition had been thought to be caused by BEC, which was
 confirmed by neutron scattering experiments\upcite{Sokol}.  
With a BEC, a macroscopic number of particles occupies a single particle ground state and is
 described by a macroscopic wave function.  
The inviscid superflow can be described by this wave function\upcite{Huang1}.  
However the relation between BEC and superfluidity is not completely understood.  
Although superfluidity and BEC are closely related to each other, one is not necessary or
 sufficient for the other.  
For example, in a two-dimensional Bose system, Kosterlitz and Thouless proved that superfluidity
 can exist even without BEC\upcite{Kosterlitz}, and the superfluidity was actually observed in
 $^4$He films\upcite{Bishop}.  
The Bose system in a random environment might be another good example for studying the relation
 between BEC and superfluidity\upcite{Trivedi,Huang2}.  
This system has received considerable attention because localization effects allow condensed
 particles to belong to the normal fluid rather than the superfluid and BEC can separate from
 the superfluid.  
The phase diagram of this system has been discussed, showing a specific non-superfluid phase.  
Thus, studies of this system can reveal the relation between BEC and superfluidity.  

Porous glass such as Vycor is often used as a random media in experimental studies.  
Vycor glass is a from 30\% to 70\% porous glass containing wormholelike pores, the characteristic
 diameters of which vary from 30\AA ~to 100\AA.  
By adjusting the pore size and the adsorbed $^4$He coverage, we can change the density of $^4$He
 and the superfluid transition.  
By using torsional oscillators, Reppy et al.\upcite{Reppy,Crooker} interesting features at various
 pore sizes of Vycor glass or coverages were observed, particularly the behavior of the superfluid
 critical temperature and the temperature dependence of the superfluid density.  
The superfluid component has two-dimensional behavior when the pore size of Vycor glass is large,
 and becomes three-dimensional as the pore size is reduced.  
The superfluid density in such porous glass is smaller than that of bulk $^4$He and its critical
 temperature decreases with the coverage.  
Below a certain coverage, the superfluid density can no longer exist, even near 0 K.  
These results show that superfluidity is broken by the random environment.  
It is also important to find out how disorder affects the BEC.  
BEC and its elementary excitation in liquid $^4$He can be observed by neutron scattering.  
Bulk $^4$He has the usual excitations such as phonons, maxons and rotons\upcite{Tilley}.  
They are collective excitations on BEC, the excitations except for phonons being absent above the
 critical temperature.  
Dimeo et al.\upcite{Dimeo} and Plantevin et al.\upcite{Plantevin} used neutron scattering and a
 torsional oscillator to measure the elementary excitations and the superfluid transition,
 respectively, of $^4$He in porous glass.  
Surprisingly, the dispersion curve in porous glass was the same as that in the bulk, which means
 that the disorder does not affect the elementary excitations.  
Furthermore, these elementary excitations were observed even above the superfluid critical
 temperature.  
Hence, the BEC might persist in the disordered system even above the superfluid critical
 temperature.  
These results have not been understood completely yet; thus, the exact relationship between BEC and
 superfluidity remains puzzling.  

This problem is also interesting from the following theoretical standpoint.  
In a Bose system confined in a random media, the long-range-order correlation due to BEC can
 compete with the disorder, so that the BEC critical temperature can be reduced.  
Huang and Meng proposed a model of the three-dimensional dilute Bose gas in a random
 potential\upcite{Huang3} that assumed a small coverage of $^4$He in Vycor glass .  
Because it is difficult to formulate the random potential for the porous glass, they used a
 delta-functional impurity potential and analyzed their model using the Bogoliubov
 transformation and taking an ensemble average.  
They found that both BEC and superfluidity are depressed by the random potential and the
 superfluidity disappears below a critical density, even at 0 K, which is qualitatively consistent
 with the observations by Reppy et al.  
They also predicted a reentrant transition at low densities; that is, the superfluid phase enters a
 normal phase again with decreasing temperatures.  
However, the random potential of their model does not include the pore size, and thus it is
 difficult to quantitatively compare to experimental results for a range of pore sizes.  
Another model is the Bose Hubbard model with the random potential.  
By considering the transfer energy, the on-site repulsion, and the random potential, Fisher et
 al. \upcite{Fisher} found that the Bose glass phase can exist with the superfluid phase and
 the Mott insulating phase. 
The Bose glass phase is similar to the Anderson insulating phase\upcite{Anderson} in metal.  
In the Bose glass phase, the condensed particles are localized and thus do not contribute to
 superfluidity.  
Thus, the Bose glass phase could influence the collective excitations even above the superfluid
 critical temperature.  
However, the theoretical excitation energy\upcite{Krauth} for the Bose glass phase disagrees with
 measurements by neutron scattering experiment\upcite{Plantevin}, so it is not yet clear whether
 the Bose glass phase has actually been detected.  
Finally, it should be noted that Huang and Meng's model cannot describe the Bose glass phase
 because the ensemble average makes the system uniform.  

Few theoretical studies of this random system are quantitatively compare to the experiment.  
Thus, in this work, we improve Huang and Meng's model\upcite{Huang3} by adding the size dependence
 of the random potential instead of using their the delta-functional potentials.  
The strength of the random potential can be estimated by comparing calculated and experimental
 critical coverages below which the superfluid density disappears, even at 0 K.  
As a result, our model has no free parameters and can be used for quantitative comparisons to
 experimental data.  
This enables us to determine whether or not our picture of the three-dimensional dilute Bose gas in
 random potential is applicable to a real system.  
Our formulation cannot address this question at high temperatures due to the high number of
 thermally excited quasi particles.  
As far as the condensate density is almost independent of temperature at low temperatures,
 however, our formulation works well, leading to the following results.  
(1) The specific heat agrees quantitatively with experimental data at low temperatures.  
(2) Because of the random potential, the specific heat is not proportional to $T^3$, as occurs for
 phonons, but to $T$.  
Furthermore, by obtaining the condensate density and the superfluid density, we found the
 following.  
(3) When the total density is sufficiently low, BEC can persist even when the superfluid density
 disappears below that critical coverage.  
(4) The random potential causes a reentrant transition of the superfluid phase.  
Finally, we show why decreasing the open pore density of the Vycor glass should allow the reentrant
 phase to be detected experimentally

A brief summary of our paper is as follows.  
In Sec. II, we describe our model of the dilute Bose gas in a random potential and derive the
 partition function.  
Section III tests our model by quantitatively comparing calculated to experimental specific
 heats.  
In Sec. IV, the BEC density and the superfluid density are obtained and their characteristics are
 discussed.  
Section V is discussion and conclusions.  

\section{Model}
Superfluid $^4$He adsorbed in Vycor glass can be modeled by a three-dimensional dilute Bose gas in
 a random external potential\upcite{Huang3}.  
The grand canonical Hamiltonian is 
\begin{eqnarray}
   \hat{H} - \mu \hat{N} &\equiv& \hat{K} \nonumber \\
     &=& \int dx^3 \hat{\Psi}^{\dagger} (\Vec{x}) \Big[ - \frac{\hbar^2}{2m}
    \Vec{\nabla}^2 + U (\Vec{x}) - \mu \Big] \hat{\Psi} (\Vec{x}) \nonumber \\
     && {} + \frac{v_0}{2} \int dx^3 \hat{\Psi}^{\dagger} (\Vec{x}) \hat{\Psi}^{\dagger} (\Vec{x})
    \hat{\Psi} (\Vec{x}) \hat{\Psi} (\Vec{x}), 
\end{eqnarray}
where $\hat{\Psi} (\Vec{x})$ is the field operator for Bose particles of mass $m$, $\hat{N} =
 \int dx^3 \hat{\Psi}^{\dagger} (\Vec{x}) \hat{\Psi} (\Vec{x})$ is the number operator, $\mu$ is
 the chemical potential,  and $U (\Vec{x})$ is the external random potential that represents the
 effect of Vycor glass.  
The first term of the Hamiltonian is the kinetic energy and the external potential, whereas the
 second term refers to the hard-sphere interaction between particles with $v_0 = 4 \pi a
 \hbar^2 / m$ being the coupling constant with the s-wave scattering length $a$.  
This repulsive interaction prevents all particles from being localized at the minimum of $U
 (\Vec{x})$.  
This has similarities to the Fermi system with disorder\upcite{Anderson,Langer,Gor'kov}; for
 example, Fermions cannot localize in a single orbital in space due to the Pauli exclusion
 principle.   
Therefore, the Fermion system is stable even if it is free from the repulsive interaction.  
On the other hand, to prevent the system from collapsing into the minimum of $U (\Vec{x})$, the
 Bose system should  include  a repulsive interaction.   
This makes the problem more complicated than that of the Fermi system.  

Proceeding in a standard fashion, we introduce the free particle annihilation and creation
 operators $\hat{a}_{\Vec{k}}$ and $\hat{a}_{\Vec{k}}^{\dagger}$.  
We assume that the level with $\Vec{k} = \Vec{0}$ is macroscopically occupied with occupation
 number $N_0$, so $\hat{a}_0$ and $\hat{a}_0^{\dagger}$ are replaced by a c-number $\sqrt{N_0}$.  
By making a Fourier transformation and neglecting all off-diagonal terms $U_{\Vec{k}}
 \hat{a}_{\Vec{k}'}^{\dagger} \hat{a}_{\Vec{k}''}$ and $v_0 \hat{a}_{\Vec{k}}^{\dagger}
 \hat{a}_{\Vec{k}'}^{\dagger} \hat{a}_{\Vec{k}''} \hat{a}_{\Vec{k}'''}$, we obtain 
\begin{widetext}
\begin{eqnarray}
   \hat{K}_{\eff} &=& V \Big( - \mu n_0 + \frac{1}{2} v_0 n_0^2 + \frac{n}{V} U_0 \Big)
    + \sum_{\Vec{k} \neq 0} \Big[ \frac{\hbar^2 k^2}{2m} - \mu + v_0 (n + n_0) \Big]
    \hat{a}_{\Vec{k}}^{\dagger} \hat{a}_{\Vec{k}} 
    + \sqrt{\frac{n_0}{V}} \sum_{\Vec{k} \neq 0}
    \big( U_{\Vec{k}} \hat{a}_{\Vec{k}}^{\dagger} + U_{-\Vec{k}} \hat{a}_{\Vec{k}} \big) \nonumber
    \\
    && {} + \frac{1}{2} v_0 n_0 \sum_{\Vec{k} \neq 0}
    \big( \hat{a}_{\Vec{k}}^{\dagger} \hat{a}_{-\Vec{k}}^{\dagger} + \hat{a}_{\Vec{k}}
    \hat{a}_{-\Vec{k}} \big), 
\end{eqnarray}
\end{widetext}
\noindent where $V$ is the volume of the system, $n_0 = N_0 / V$ is the number density of
 condensate, and $U_{\Vec{k}}$ is the Fourier transformation of $U(\Vec{x})$.  
By neglecting the off diagonal terms, we are neglecting the interactions between the excited
 particle and the random potential and that between pairs of excited particles; these become
 important as the temperature rises and the condensate density decreases.  
Hence, this approximation is poor when many particles are thermally excited.  
Nevertheless, this approximation is useful at low temperatures where the condensate density is
 almost independent of temperature.  
All results here are obtained for these low temperatures.  

This Hamiltonian can be diagonalized by the Bogoliubov transformation 
\begin{equation}
   \hat{a}_{\Vec{k}} = \frac{\hat{c}_{\Vec{k}} + \gamma_{\Vec{k}} \hat{c}_{-\Vec{k}}^{\dagger}}
   {\sqrt{1-\gamma_{\Vec{k}}^2}} + g_{\Vec{k}}.  \label{eq:Bogoliubov}
\end{equation}
Then the coefficients $\gamma_{\Vec{k}}$, $g_{\Vec{k}}$ and the quasiparticle spectrum
 $\omega_{\Vec{k}}$ are given by
\begin{subequations}
\begin{eqnarray}
    \gamma_{\Vec{k}} &=& - \xi - 1 + \sqrt{\xi (\xi + 2)}, \\
    g_{\Vec{k}} &=& - \sqrt{\frac{n_0}{V}} \frac{U_{\Vec{k}}}{(\xi + 2) v_0 n_0}, \\ 
    \omega_{\Vec{k}} &=& v_0 n_0 \sqrt{\xi (\xi + 2)}, \\
    \xi &=& \frac{\hbar^2 k^2}{2 m v_0 n_0} + \Delta, \quad \Delta = \frac{v_0 n_0 - \mu}{v_0 n_0}
    \label{eq:gapspectrum}.  
\end{eqnarray}
\end{subequations}

Next, we take an ensemble-average to quench the random potential.  
The random potential simulates Vycor glass with a characteristic pore size $r_p$ as follows.  
The quenched potential $U_{\Vec{k}}$ may decay above the characteristic wave number $k_p = 2 \pi
 / r_p$.  
Thus, we assume the averaged potential
\begin{equation}
   \frac{1}{V} \big\langle U_{\Vec{k}} U_{-\Vec{k}} \big\rangle_{\av} = R_0
    \exp\Big[ - \frac{k^2}{2 k_p ^2} \Big], \label{eq:ensemble-av}
\end{equation}
where av denotes the ensemble-average.  
$R_0$, with dimension $(energy)^2 (length)^3$, is the characteristic strength of the random
 potential.  
Equation (\ref{eq:ensemble-av}) makes our model completely different from Huang and Meng's, and we
 will show that the results are also different.  
The coherence length of the BEC is thought to be from hundreds to thousands of \AA, whereas the
 spatial scale of disorder is the pore size in the glass, which is dozens of \AA.  
Hence, the macroscopic wave function of BEC is not sensitive to disorder in and between pores but
 instead depend on the disorder averaged over the coherence length.  
Hence, the ensemble-averaged system can become nearly uniform.  
For a uniform Bose system, it has been proven that the elementary excitation spectrum becomes the gapless
 Goldstone mode\upcite{Stringari}.  
Thus, we set $\Delta = 0$ in Eq. (\ref{eq:gapspectrum}).  

The resultant diagonalized and ensemble-averaged Hamiltonian is 
\begin{subequations}
\begin{eqnarray}
   \hat{K}_{\eff} &=& V( - \mu n_0 + \frac{n}{V} U_0 + \epsilon_1 + \epsilon_R ) \nonumber \\
   && {} + \sum_{\Vec{k} \neq 0} \hbar \omega_k \hat{c}_{\Vec{k}}^{\dagger} \hat{c}_{\Vec{k}}, 
\end{eqnarray}
\begin{eqnarray}
    \omega_k &=& \frac{\hbar}{2m} k \sqrt{k^2 + 16 \pi a n_0}, \\
    \epsilon_1 &=& \frac{2 \pi a n_0^2 \hbar^2}{m} \Big[ 1 + \frac{128}{15 \sqrt{\pi}}
    \sqrt{n_0 a^3} \Big], \\
    \epsilon_R &=& \frac{m \sqrt{a} n_0^{3/2} R_0}{4 \sqrt{\pi} \hbar^2} \Big[ - \e^{2 \alpha} (5
    + 4 \alpha) \{1 - \erf(\sqrt{2 \alpha}\} \nonumber \\
    && \hspace{21mm} + \sqrt{\frac{2}{\pi \alpha}} (1 + \alpha) \Big], \\
    \alpha &=& \frac{4 \pi a n_0}{k_p^2}, 
\end{eqnarray}
\end{subequations}
where $\epsilon_1$ is the hard sphere interaction energy at 0 K, similarly, $\epsilon_R$ is that
 for the random potential.  
The quasiparticle spectrum $\omega_k$ is the same as that in the hard sphere Bose gas
 model\upcite{Huang2} and is independent of the random potential.  
This independence is confirmed by neutron scattering experiments\upcite{Plantevin}, which
 justifies the above assumption of $\Delta = 0$; conversely, if $\Delta \neq 0$, the spectrum would
 depend on the random potential.  

This Hamiltonian enables us to obtain the grand partition function $Q = \Tr \{ \exp ( - \beta
 \hat{K} ) \}$ and various physical quantities.
The condensate density is defined by the following relation:
\begin{equation}
   n_0 = n - \frac{1}{V} \sum_{\Vec{k} \neq 0} \big\langle \hat{a}_{\Vec{k}}^{\dagger}
    \hat{a}_{\Vec{k}} \big\rangle \label{eq:condpart}, 
\end{equation}
where $n$ is the particle number density.  
The second term represents the noncondensate particle number as
\begin{subequations}
\begin{eqnarray}
   \frac{1}{V} \sum_{\Vec{k} \neq 0} \big\langle \hat{a}_{\Vec{k}}^{\dagger} \hat{a}_{\Vec{k}}
    \big\rangle = n_1 + n_R, 
\end{eqnarray}
\begin{eqnarray}
    n_1 &=& \frac{8}{3 \sqrt{\pi}} (n_0 a)^{3/2} \nonumber \\
    && {} + \frac{4}{\sqrt{\pi} {\lambda}^3} \int_0^\infty dt \frac{t (t^2 + {\theta} /
    2)}{\sqrt{t^2 + \theta} \{ \e^{t \sqrt{t^2 + \theta}} - 1 \}}, \\
    n_R &=& \frac{m^2 R_0}{8 \pi^{3 / 2} \hbar^4} \sqrt{\frac{n_0}{a}} \Big[ \e^{2 \alpha} (1
    + 4 \alpha) \{1 - \erf(\sqrt{2 \alpha})\} \nonumber \\
    && \hspace{23mm} - 2 \sqrt{\frac{2 \alpha}{\pi}} \Big], \\
    \lambda &=& \sqrt{\frac{2 \pi \beta \hbar^2}{m}}, \quad \theta =
    \frac{8 \pi a \hbar^2 \beta n_0}{m},
     \quad t^2 = \frac{\hbar^2 \beta}{2 m} k^2.  
\end{eqnarray}
\label{eq:realcond}
\end{subequations}
Here $n_1$ is the noncondensate density excited by the hard sphere interaction, $n_R$ is the
 density due to the scattering of condensate particles with the random potential, and $\lambda$ is
 the thermal de Broglie wave length.  
When $a$ vanishes, $n_R$ becomes infinite.  
This means that the system would collapse if there were no repulsive interactions between
 particles.  

Because superfluidity is described by the two fluid model, the particle density $n$ consists of the
 normal fluid density $n_n$ and the superfluid density $n_s$.  
The superfluid density $n_s$ can be calculated by linear response theory\upcite{Hohenberg}.  
Because of its viscosity, only the normal fluid responds to a small, applied velocity field.  
Thus the normal fluid density can be defined by the response of the momentum density $j_i
 (\Vec{x},t)$ to the external velocity field $v_i (\Vec{x},t)$.  
Linear response theory gives the following relation:
\begin{subequations}
\begin{eqnarray}
    j_i (\Vec{x},t) &=& \chi_{ij} (\Vec{x},t) v_j (\Vec{x},t), \\
    \chi_{ij} (\Vec{x},t) &=& \langle [ j_i (\Vec{x},t),j_j (0,0) ] \rangle, \\
    j_i(\Vec{x},t) &=& \frac{\hbar}{2i} \Big\{ \hat{\Psi}^{\dagger} (\Vec{x},t) \frac{\partial
    \hat{\Psi} (\Vec{x},t)}{\partial x_i} \nonumber \\
    && \hspace{7mm} - \frac{\partial \hat{\Psi}^{\dagger} (\Vec{x},t)} {\partial x_i} \hat{\Psi}
    (\Vec{x},t) \Big\}, \\
    \hat{\Psi} (\Vec{x},t) &=& \e^{ \mathrm{i} (\hat{H} - \mu \hat{N}) t / \hbar } \hat{\Psi}
    (\Vec{x}) \e^{ - \mathrm{i} (\hat{H} - \mu \hat{N}) t / \hbar },  
\end{eqnarray}
\end{subequations}
\noindent where $\hat{\Psi} (\Vec{x},t)$ is the Heisenberg field operator.  
The static susceptibility $\chi_{ij} (\Vec{k})$ is defined as
\begin{subequations}
\begin{eqnarray}
   \chi_{ij} (\Vec{x},t) &=& \int \frac{d \omega}{2 \pi} \frac{d^3 k}{(2 \pi)^3} \e^{- \mathrm{i}
    \omega t} \e^{\mathrm{i} \Vec{k} \cdot \Vec{x}} \chi_{ij} (\Vec{k},\omega), \\
   \chi_{ij} (\Vec{k}) &=& \lim_{\omega \to 0} \chi_{ij} (\Vec{k},\omega).  
\end{eqnarray}
\end{subequations}
Because of the rotational invariance, the static susceptibility $\chi_{ij} (\Vec{k})$ can be
 written
\begin{equation}
   \chi_{ij} (\Vec{k}) = \frac{k_i k_j}{k^2} A(\Vec{k}) + \Big( \delta_{ij} - \frac{k_i k_j}{k^2}
    \Big) B(\Vec{k}), 
\end{equation}
where $A(\Vec{k})$ and $B(\Vec{k})$ are the longitudinal and transverse parts, respectively.  
The transverse susceptibility $B(0)$ is the normal fluid mass density.  
The superfluid number density $n_s$ is $n - B(0)/m$.  
The susceptibility $B(0)$ can be calculated by the Bogoliubov transformation in
 Eq. (\ref{eq:Bogoliubov}).  
After some tedious calculations, the resultant superfluid density is given by
\begin{subequations}
\begin{eqnarray}
   && n_s = n - n_{n1} - n_{nR}, \\
   && n_{n1} = \frac{8}{3 \sqrt{\pi} \lambda^3} \int_0^\infty dt \frac{t^4 \e^{ - t \sqrt{t^2
    + \theta} }}{ ( 1 - \e^{ - t \sqrt{t^2 + \theta} } )^2 }, \\
   && n_{nR} = \frac{4}{3} n_R, 
\end{eqnarray}
\label{eq:super}
\end{subequations}
where $n_{n1}$ is the normal fluid density due to the elementary excitations, and $n_{nR}$ is that
 due to scattering with the random potential.  
The density $n_{n1}$ can be also obtained using Khalatnikov's method that is based on Galilean
 invariance\upcite{Khalatnikov}.  
The relation $n_{nR} = 4 / 3 n_R = n_R + 1 / 3 n_R$ shows that the random potential causes the
 larger normal fluid density than the noncondensate density; some condensate particles are captured
 by the random potential to participate in the normal fluid.  
This makes it possible to destroy superfluidity even at 0 K when $n_{nR}$ becomes comparable to
 $n$.  

This formulation can be used to obtain various physical quantities including the condensate
 density, the superfluid density, and the specific heat.  

\section{Comparison with experiments}
In this section, we compare the calculated specific heat and the superfluid density with
 experimental results.  
Quantitative agreement is shown to be good at low temperatures, which supports our assumption of a
 dilute Bose gas in the random potential.  
Furthermore, we show that the random potential leads to as-yet unobserved behavior of the
 specific heat.  

\begin{figure}[htbp]
\begin{minipage}{0.9\linewidth}
\begin{center}
\includegraphics[width=0.9\linewidth]{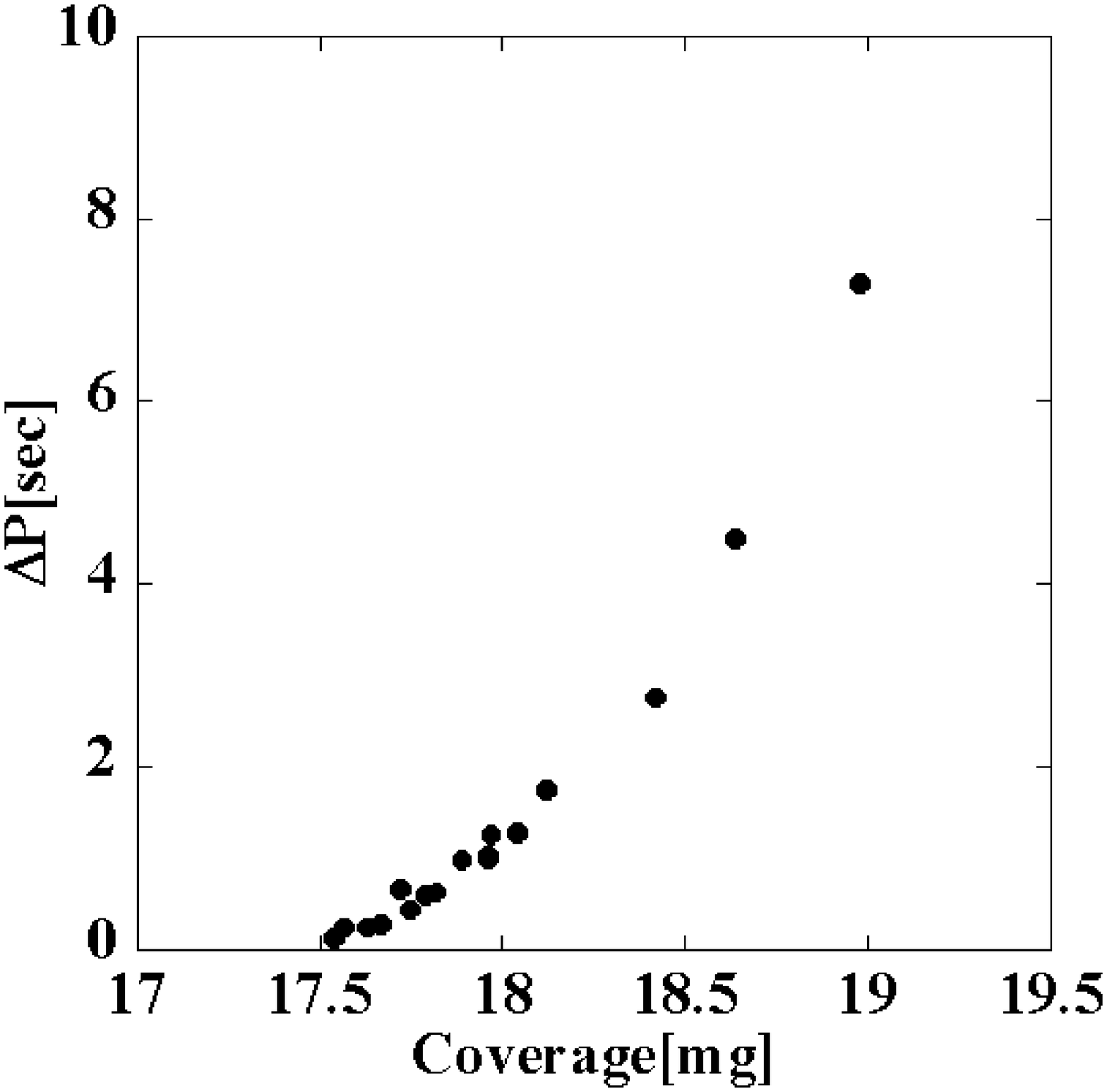} \\
(a)
\end{center}
\end{minipage}

\begin{minipage}{0.9\linewidth}
\begin{center}
\includegraphics[width=0.9\linewidth]{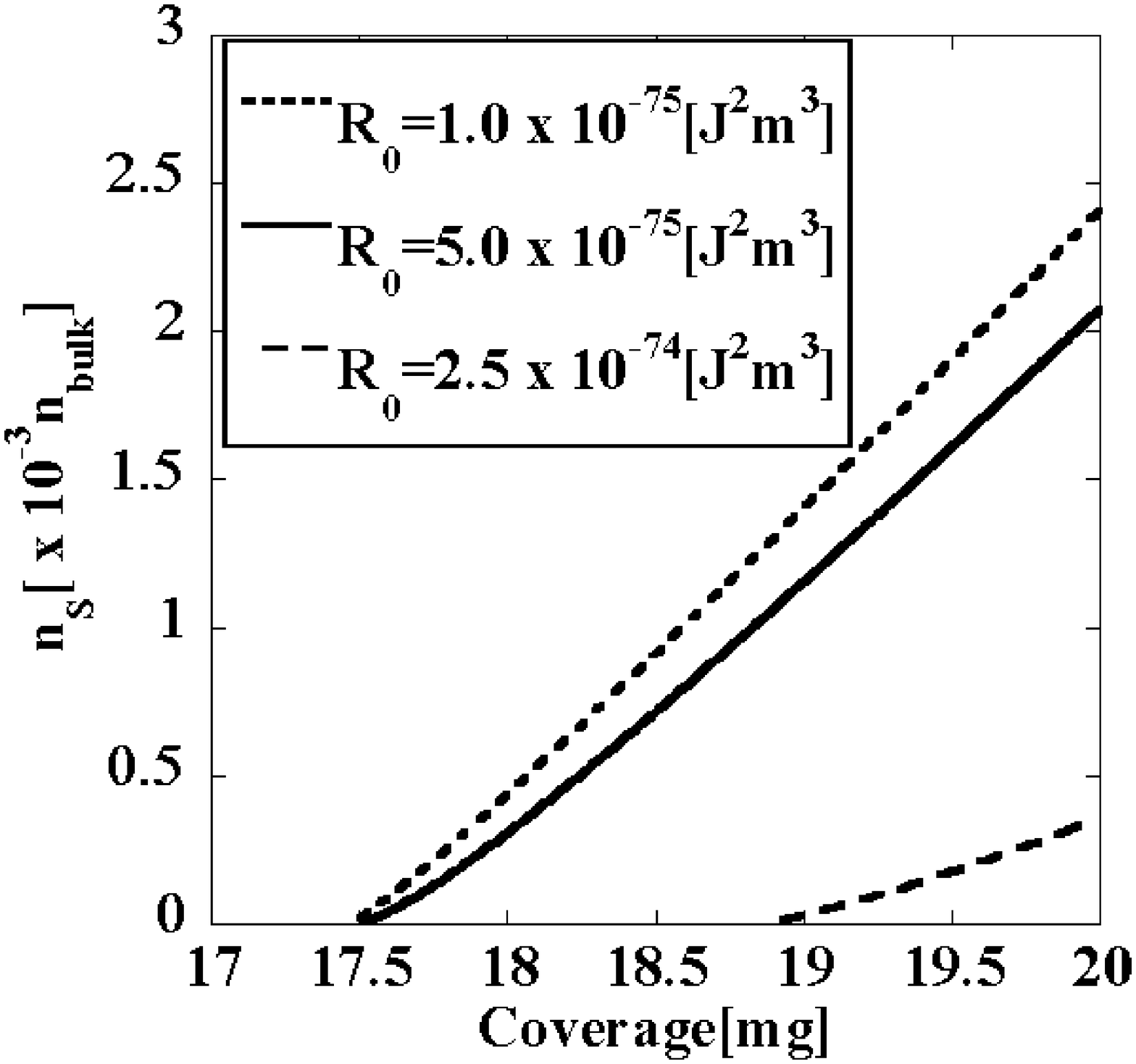} \\
(b)
\end{center}
\end{minipage}
\caption{Superfluid signals of experiments (a) and calculations (b) near 0 K.  
$\Delta P$ is the resonant period difference in a torsional balance experiment, which is
 approximately proportional to the superfluid density\upcite{Reppy}.  }
\label{zerosuper}
\end{figure}

To make a quantitative comparison, we give the following numerical values to the parameters: $m
 \simeq 6.6 \times 10^{-27} ~\kg$ and $a \simeq 5 \times 10^{-10} ~\m$ are the mass and the s-wave
 scattering length of a $^4$He atom.  
Other parameters are from the experiments of Reppy et al.\upcite{Reppy}.  
The volume $V$ of open pores in the Vycor glass (about 40\% of the total volume of the Vycor glass)
 is about 1 $\cm^3$.  
The particle density $n$ of $^4$He inside the Vycor glass is estimated as follows.  
In Vycor glass, the atoms are adsorbed and fully cover the surfaces of the open pores due to the
 van der waals attraction.  
The pore area is about $10^8$ $\m^2/\m^3$.  
The rest of atoms, which do not participate in the first-layer solid, can behave as a dilute gas
 inside the pores.  
The particle density $n$ of the dilute gas is obtained by subtracting the adsorbed amount from the
 total amount.  
This density is estimated to be from 0.001\% to 70\% of the density of bulk liquid $^4$He $n_{bulk}
 \sim 2.1 \times 10^{28} / \m^3$.  
Because the first layer of $^4$He adsorbed on the surfaces cannot move and behaves as a solid, we
 assume the pore size 30\AA~of Vycor glass is effectively reduced by $2 a$.  
Thus, $r_p$ is estimated to be 20\AA.
The last parameter $R_0$, which is the strength of the random potential, can be fixed by comparing
 to experiment.  
Shown in Fig.~\ref{zerosuper}(a) are the data of zero temperature superfluid signals taken in an
 experiment that used a torsional oscillator (Fig.~12 of reference\upcite{Reppy}).  
Because the superfluid component does not contribute to the moment of inertia, the resonant
 frequency and the period of oscillation differ from those without superfluid.  
The period difference $\Delta P$ is approximately proportional to the superfluid component.  
Here, the superfluid density is nearly proportional to $\Delta P$ and disappears at a coverage of
 17.5 mg.  
Figure~\ref{zerosuper}(b) shows the superfluid density at 0 K from Eq. (\ref{eq:super}).  
As in the experiment, the superfluid density becomes zero at a certain coverage that depends on
 $R_0$.  
Thus, the value of $R_0$ can be fixed using the comparison with Fig.~\ref{zerosuper}(a); i.e., $R_0
 = 5 \times 10^{-75}~\J^2 \m^3$.  
Here, we define $R_w \equiv \sqrt{R_0 n}$, which is the single particle energy converted from
 $R_0$.  
In the Vycor glass, $R_w / k_B$ is about $0.001 \sim 1$ K.  
Just above the critical coverage, the superfluid density increases linearly for both the experiment
 and the calculation; however, their slopes cannot be compared because the amplitude of $n_s$ is
 unknown in the experiment.  

\begin{figure}[htbp]
\begin{minipage}{0.9\linewidth}
\begin{center}
\includegraphics[width=0.9\linewidth]{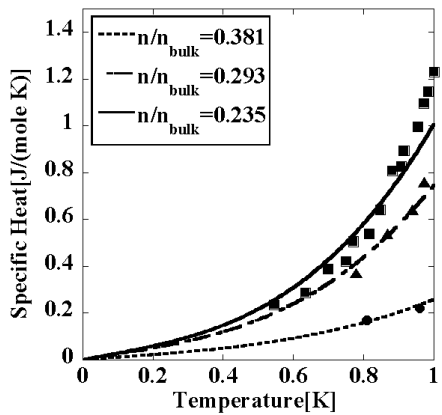} \\
(a)
\end{center}
\end{minipage}

\begin{minipage}{0.9\linewidth}
\begin{center}
\includegraphics[width=0.9\linewidth]{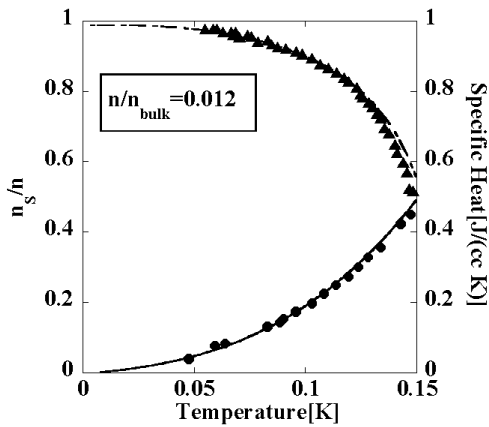} \\
(b)
\end{center}
\end{minipage}
\caption{The specific heat data from in experiments (plot) and calculations (line).  
In (a), experimental data are given by Fig.1 of reference\upcite{Reppy}.  
The circles, triangles and squares respectively correspond to full pores ($\sigma = 1$), $\sigma =
 0.780$ and $\sigma = 0.636$.  
Here $\sigma$ is the ratio of the coverage to the full pore coverage.  
In (b), calculated and experimental superfluid densities are compared ($T_c = 0.163 \K$).  }
\label{speheat}
\end{figure}

Because all parameters are now fixed, we will quantitatively compare calculations to experiments.  
The specific heat can be obtained from temperature differentiation of the free energy: 
\begin{subequations}
\begin{eqnarray}
    \Omega &=& \Omega_1 + \Omega_R, \\
    \Omega_1 &=& V( - \mu n_0 + \epsilon_1) \nonumber \\
    && {} + \frac{4 V}{\sqrt{\pi} \beta \lambda^3} \int_0^\infty dt \{ t^2 \log[1 -
    \e^{- t \sqrt{t^2 + \theta}}] \}, \\
    \Omega_R &=& V( \epsilon_R + \frac{n}{V} U_0).  
\end{eqnarray}
\label{eq:specific}
\end{subequations}
where $\Omega_1$ is the free energy of the elementary excitation and the hard sphere interaction, 
and $\Omega_R$ is the free energy from the random potential.  
Figure~\ref{speheat} compares our results to data of low temperature specific heat taken from
 Fig.~1 of reference\upcite{Reppy}.  
In Fig.~\ref{speheat}(a), which shows the data for high density, the density $n$ is fixed from
 the experimental coverage, whereas we fix the density from the superfluid critical temperature in
 Fig.~\ref{speheat}(b) at low density.  
This is because we have no information about the data on the coverage.  
Figure~\ref{speheat}(b) also shows the superfluid density.  
The theoretical results agree quantitatively with experiment without using free parameters.  
Above 1.0 K in Fig.~\ref{speheat}(a), the calculated condensate density begins to decrease
 rapidly; here, our criterion of constant condensate density fails, which likely causes the
 discrepancy with experiment.  
However, Fig.~\ref{speheat}(b) shows that the calculated specific heat agrees with experiment up to
 temperatures near the superfluid critical temperature; in this temperature region, the calculated
 condensate density hardly decreases.  
This means that the system is more dilute than that of Fig.~\ref{speheat}(a) and thus is affected
 by the random potential rather than the elementary excitations.  
We discuss this in the next section.  
Nevertheless, these comparisons show that our model is accurate at low temperatures.  

\begin{figure}[htbp]
\includegraphics[width=0.9\linewidth]{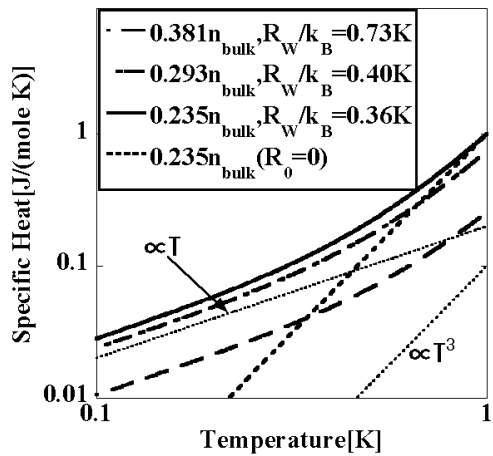}
\caption{The log-log plot of Fig.~\ref{speheat}(a).  
Data of the specific heat at $R_0 = 0 (n/n_{bulk} = 0.35)$.  
Two lines for $\propto T$ and $\propto T^3$ are added.  }
\label{logspe}
\end{figure}

Our model predicts an effect at low temperature that is due to the random potential but has not yet
 been observed.  
This is shown in Fig.~\ref{logspe}, which is the log-log plot of Fig.~\ref{speheat}(a).  
When the system is free from the random potential, the specific heat should increase a $T^3$
 because of the contribution from the phonons.  
However, with a random potential, the dependence is linear in $T$ at low temperatures.  
This means that at low temperatures, the contribution from the random potential is larger than that
 from the elementary excitations (phonons).
The free energy $\Omega_R$ in Eq. (\ref{eq:specific}) depends on the temperature only through the
 condensate density $n_0$.  
$\Omega_R$ is the energy from the scattering of the condensate particles with the random potential,
 and the resultant specific heat is given by the energy that the condensate particles need to slip
 out of the random potential.  
Experimental observation of this $T$-linear dependence might clearly identify the influence from
 the random potential.  

\section{Condensate density and superfluid density}
This section describes some characteristic behavior of the condensate density $n_0$ and the
 superfluid density $n_s$ derived from our model.  

\begin{figure}[htbp]
\begin{minipage}{1.0\linewidth}
\begin{center}
\includegraphics[height=0.28\textheight]{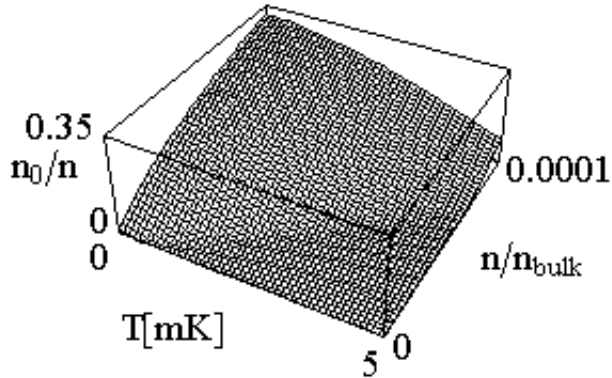} \\
(a)
\end{center}
\end{minipage}

\begin{minipage}{1.0\linewidth}
\begin{center}
\includegraphics[height=0.28\textheight]{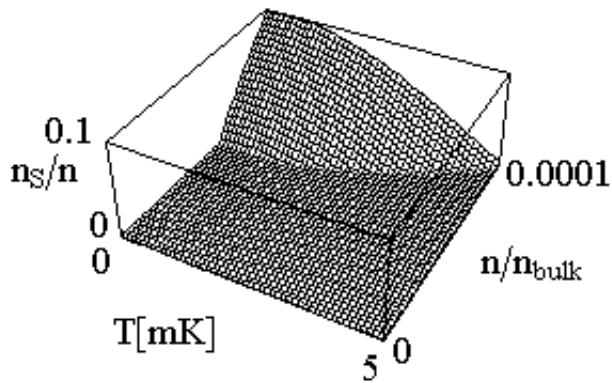} \\
(b)
\end{center}
\end{minipage}
\caption{Temperature and density dependence of $n_0$ (a) and $n_s$ (b) at $R_0 = 5.0 \times
 10^{- 75} \J^2 \m^3$.  }
\label{3D}
\end{figure}

Figure.~\ref{3D} shows the dependence of $n_0$ and $n_s$ on temperature and density.  
Both $n_0$ and $n_s$ decrease with decreasing density, even at 0 K.  
This means that the effect of the random potential on $n_0$ and $n_s$ becomes larger as the density
 is reduced.  
Figure.~\ref{3D} shows clearly the difference between $n_s$ and $n_0$.  
Below the critical density, the superfluid density disappears, although the condensate persists.  
This situation indicates that the condensate particles cannot move as a superfluid because they are
 trapped by the random potential.  
We expect that this theoretical result will be confirmed by measurements of condensate density.  

\begin{figure}[htbp]
\includegraphics[width=0.9\linewidth]{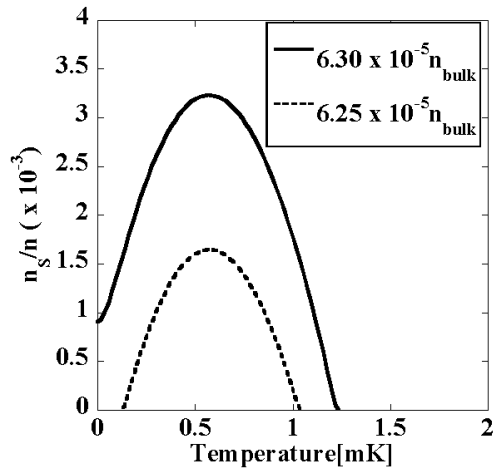}
\caption{Temperature dependence of $n_s$ at low temperature and low density.  
In this regime, the superfluid density $n_s$ goes to zero with a decrease of temperature (reentrant
 transition).  }
\label{reent}
\end{figure}

\begin{figure}[htbp]
\includegraphics[width=0.9\linewidth]{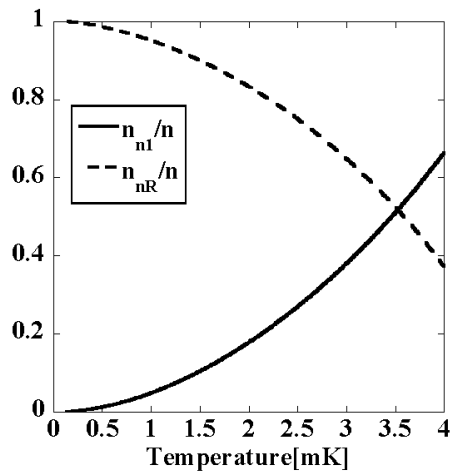}
\caption{Temperature dependence of $n_{n1}$ and $n_{nR}$ near the reentrant distribution at 
$n = 6.25 \times 10^{-5}~n_{bulk}$ (dashed line in Fig.~\ref{reent}).  }
\label{reentreason}
\end{figure}

\begin{figure}[htbp]
\begin{minipage}{0.9\linewidth}
\begin{center}
\includegraphics[width=0.9\linewidth]{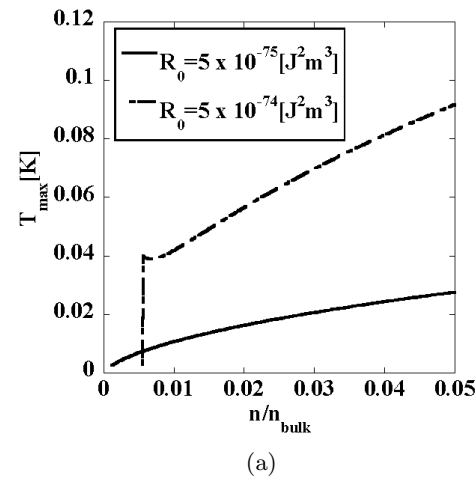} \\
(a)
\end{center}
\end{minipage}

\begin{minipage}{0.9\linewidth}
\begin{center}
\includegraphics[width=0.9\linewidth]{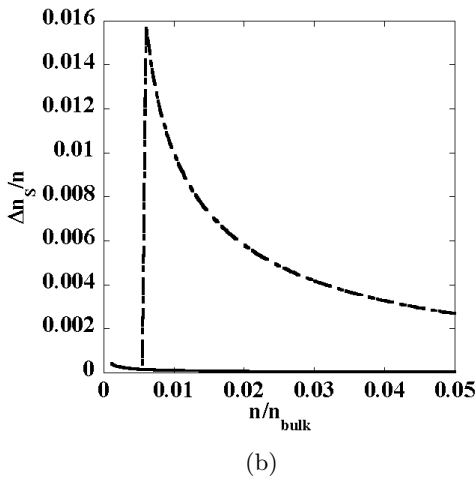} \\
(b)
\end{center}
\end{minipage}
\caption{Density dependence of $T_{max}$ and $\Delta n_s$ for two values of $R_0$.  
$T_{max}$ is the temperature that maximizes the superfluid density $n_s$, and $\Delta n_s = n_s(T =
 T_{max}) - n_s(T = 0)$.  }
\label{reentreason2}
\end{figure}

Figure.~\ref{reent} shows the temperature dependence of the superfluid density just before
 superfluidity disappears.  
This figure shows the reentrant transition at which the superfluid density $n_s$ goes to zero with
 a decrease of temperature.  
In this temperature region, the condensate density is almost constant; hence, our formulation
 should work well in accordance with the criterion described in Sec. II.  
This reentrant transition is understood as follows.  
The condensate depletion $n_R$ of Eq.~(\ref{eq:realcond}c) comes from the scattering of the
 condensate particles with the random potential.  
This decreases with $n_0$ as the temperature rises, so that $n_{nR} = 4 / 3 n_R$ of
 Eq.~(\ref{eq:super}c) also decreases.  
As shown in Fig.~\ref{reentreason}, the magnitude of this decrease in $n_{nR}$ exceeds the increase
 in $n_{n1}$ of Eq.~(\ref{eq:super}b), which is the normal fluid density due to the elementary
 excitations, in the very low temperature region $T \le 0.5$ mK.  
In other words, condensate particles that are trapped by the random potential at lower temperatures
 can escape at higher temperatures and thus participate in superfluidity.  
This reentrant transition has not been observed experimentally, probably because it should only
 occur at very low densities and low temperatures.  
However, large values of $R_0$ can make the reentrant transition observable as follows.  
We define the temperature $T_{max}$ as that which maximizes the superfluid density $n_s$, and
 define $\Delta n_s \equiv n_s(T = T_{max}) - n_s(T = 0)$.  
Fig.~\ref{reentreason2} shows the density dependence of $T_{max}$ and $\Delta n_s$.  
Both variables increase with $R_0$.  
Therefore, the reentrant transition is more likely to be measured at larger $R_0$.  
The parameter $R_0$ is the strength of the random potential over the entire space, and one way to
 increase $R_0$ is to decrease the open pore density of the Vycor glass.  

\section{Conclusions}
The present paper describes the dilute Bose gas system in a random potential.  
The outcomes of our studies are as follows.  

By including the pore size dependence of Vycor glass in the random potential, our model could
 closely match the experimental conditions of liquid $^4$He in Vycor glass.  
We fixed the strength of the random potential by equating the theoretical and experimental critical
 coverages below which the superfluid density at 0 K vanishes.  
No other parameters could be adjusted, and thus we could quantitatively compare theory to
 experiment for other physical quantities.  

First, we showed that the calculated specific heat for Vycor glass quantitatively agrees with
 measurements.  
This indicates that liquid $^4$He in Vycor glass behaves as a dilute Bose gas in a random
 potential.  
For low temperatures, the calculated specific heat was linear in $T$ because of the random
 potential.  
Second, the BEC was shown to persist even when superfluidity disappears below the critical
 density.  
Finally, we showed that the reentrant transition of the superfluid phase is more likely to be
 observed experimentally by increasing the strength of the random potential.  

Because we neglected interactions between pairs of excited particles and between excited particles
 and the random potential, this model does not apply to systems at high temperatures.  
To overcome this limitation, we are improving the model to include these interactions and will
 report on this more general model in the near future.  

\begin{acknowledgments}
MT acknowledges support by a Grant-in-Aid for Scientific Research (Grant No.12640357) by Japan
 Society for the Promotion of Science.  
\end{acknowledgments}

\end{document}